\documentclass[reprint,
twocolum,
groupedaddress,
showpacs,
 amsmath,amssymb,
 aps,
prb,
]{revtex4-1}

\usepackage{graphicx}
\usepackage{wrapfig}
\usepackage{dcolumn}
\usepackage{bm}
\usepackage{epstopdf}
\usepackage{amsmath}              
\usepackage{amssymb} 
\usepackage[justification=raggedright]{caption}
\usepackage{textcomp}
\usepackage{tabularx} 
\usepackage{multirow}
\usepackage[ngerman, english]{babel}

\usepackage{hyperref}

\begin{document}
\preprint{}

\title{Electronic phase diagram of disordered Co doped BaFe\texorpdfstring{$_2$}{2}As\texorpdfstring{$_2$}{2}}

\author{F.~Kurth$^{1,2}$}
\email[]{Fritz.Kurth@ifw-dresden.de}
\author{K.~Iida$^1$, S.~Trommler$^{1,2}$, J.~H\"anisch$^1$, K.~Nenkov$^1$, J.~Engelmann$^{1,2}$, S.~Oswald$^1$, J.~Werner$^1$, L.~Schultz$^{1,2}$, B.~Holzapfel$^{1,3}$, S.~Haindl$^{1}$}
\email[]{S.Haindl@ifw-dresden.de}
\affiliation{$^1$IFW Dresden, Helmholtzstr.\ 20, 01069 Dresden, Germany\\
$^2$TU Dresden, 01062 Dresden, Germany\\
$^3$TU Bergakademie Freiberg, Akademiestr.\ 6, 09596 Freiberg, Germany
}

\begin{abstract}
Superconducting and normal state transport properties in iron pnictides are sensitive to disorder and impurity scattering. By investigation of Ba(Fe$_{1-x}$Co$_x$)$_2$As$_2$ thin films with varying Co concentration, we demonstrate that in the dirty limit the superconducting dome in the electronic phase diagram of Ba(Fe$_{1-x}$Co$_x$)$_2$As$_2$ shifts towards lower doping concentrations, which differs significantly from observations in single crystals. We show that especially in the underdoped regime superconducting transition temperatures higher than 27~K are possible.

\begin{description}
\item[PACS numbers]
\verb+{74.70.-b, 74.62.-c, 74.25.-q}+
\end{description}
\end{abstract}

\pacs{74.70.-b, 74.62.-c, 74.25.-q}
\maketitle


\section{\label{sec:level1}Introduction\protect}
Almost simultaneously with the report of superconductivity by hole doping in BaFe$_2$As$_2$,\cite{PhysRevLett.101.107006} Sefat {\itshape et al.} showed that also electron doping in BaFe$_2$As$_2$ is able to supress the spin density wave (SDW) state and to stabilize superconductivity.\cite{sef01} Since then, many studies have been devoted to an understanding of the competition between the SDW and the superconducting state in this intermetallic compound.\cite{ni08, chu10, marsik10, ning09, ning10, fang2009}\\
\indent Investigations of the electronic phase diagram of Co doped BaFe$_2$As$_2$ revealed a structural transition from a high-temperature tetragonal to a low-temperature orthorhombic phase closely above the transition from the paramagnetic to the SDW state.\\
\indent With increasing doping level {\textit x} in Ba(Fe$_{1-{\textit x}}$Co$_{\textit x}$)$_2$As$_2$ the structural and magnetic transitions split further and are suppressed.\cite{ni08, chu10} In addition, a characteristic superconducting dome raises with a maximal critical temperature up to {\textit T$_\mathrm{c}$} = 25~K around the optimal doping of {\textit x} = 0.06 - 0.08 until superconductivity vanishes beyond {\textit x} = 0.18 - 0.20. In the underdoped region ({\textit x} $\le$ 0.06) superconductivity coexists with the SDW state microscopically. This extremely close vicinity of superconductivity to antiferromagnetism suggests the mediating role of spin fluctuations in the Cooper pairing.\\
\indent The influence of increasing Co content in Ba(Fe$_{1-x}$Co$_x$)$_2$As$_2$ on the electronic structure results in an increase of its three-dimensionality, a gradual filling of the hole pockets near the $\Gamma$ point of the Brillouin zone, and therefore a reduced nesting.\cite{PhysRevB.81.104512} It has been argued that the loss of nesting ({\itshape i.e.}\ the weakening of interband transitions) and the suppressed spin fluctuations are responsible for the decrease in {\textit T$_\mathrm{c}$} in overdoped samples, whereas in underdoped samples superconductivity has to compete with the SDW state for the carrier density.\cite{ning10, fang2009, PhysRevLett.103.087002, pratt09}\\
\indent A more explicit difference between the effects of electron doping on the structural, magnetic and superconducting transitions was made by Canfield {\itshape et al.}\ by comparing Co doping with Ni and Cu doping in BaFe$_2$As$_2$.\cite{canfield09} They concluded that the occurrence of superconductivity depends primarily on the amount of added electrons and the {\itshape c/a} ratio, whereas the structural modifications, defects and magnetic phase transitions depend on the amount of added impurity ions and the {\textit c}-axis parameter. In this regard other pertubations in the FeAs tetrahedral sublattice (vacancies and interstitials included) should be considered in more detail.\\
\indent All the investigations mentioned above have been performed on powder samples and on single crystals. Here, we will discuss the peculiarities in the growth of Co doped BaFe$_2$As$_2$ thin films with varying Co content and the influence on the electronic phase diagram. Since the first growth of Co doped BaFe$_2$As$_2$ thin films by pulsed laser deposition (PLD),\cite{iida:192501, lee:212505, Katase20092121} a detailed analysis of the chemical composition and homogeneity has been widely neglected. Therefore, a part of our study is devoted to the investigation of the thin film stoichiometry. Auger electron spectroscopy (AES) demonstrates that the chemical composition of the films is sensitive to the deposition conditions. One difference between thin films and single crystals of iron pnictide superconductors is their degree of disorder, easily demonstrated by RRR values. As observed in comparable films without Fe buffer, typical RRR values for Co doped BaFe$_2$As$_2$ are in the range of 1 to 2. \\

\section{\label{sec:level2}Sample Preparation}

For the preparation of the target materials the starting elements (Ba, Fe, Co and As) are mixed and mechanically milled resulting in precursor powders of Fe$_2$As, Co$_2$As, and BaAs. Precursor powders of {\textit x} = 0, 0.02, 0.04, 0.06, 0.08, 0.10, and 0.15 were synthesized and pressed in pellets of Ba(Fe$_{1-x}$Co$_x$)$_2$As$_2$ with different Co content. The oxygen impurities of the starting elements were taken into account during the weighing process. Impurities in As were negligible since As was cleaned by sublimation.\cite{Briehl1995} More details of the target preparation are given in appendix A.\\ 
\indent Using the sintered Ba(Fe$_{1-x}$Co$_x$)$_2$As$_2$ targets, thin films with different Co concentration were fabricated by PLD under ultra high vacuum (UHV) conditions (base pressure of 10$^{-9}$ mbar) on MgO(100) substrates. A KrF exciplex laser ($\lambda$ = 248~nm, pulse duration = 25~ns) was used for material ablation.\\ 
\indent Before each film deposition the target surface was cleaned by approximately 1000 laser pulses. The substrates were cleaned in an ultrasonic bath using aceton and isopropanol and subsequently transferred into the UHV chamber. A heat treatment of the substrate at 1000~\textcelsius\ for 30 minutes followed prior to the deposition process.\\ 
\indent Thin film growth started with the deposition of a pure Fe buffer at room temperature. The Ba(Fe$_{1-x}$Co$_x$)$_2$As$_2$ deposition followed at substrate temperatures between 650~\textcelsius\ and 800~\textcelsius\ depending on the nominal Co content (Tab.~\ref{tab:depbedingungen}). The deposition temperature was increased with increasing Co doping. The repetition rate was set to 5~Hz for Fe and to 10~Hz for Ba(Fe$_{1-x}$Co$_x$)$_2$As$_2$. Pulse numbers for the Fe buffer layer were set to 4000 - 5000 and for Ba(Fe$_{1-x}$Co$_x$)$_2$As$_2$ to 8000. The estimated energy densities at the target surface of Ba(Fe$_{1-x}$Co$_x$)$_2$As$_2$ are given in Tab. \ref{tab:depbedingungen}.

\begin{table*}[htbp]
\caption{Deposition conditions and structural data of Ba(Fe$_{1-x}$Co$_x$)$_2$As$_2$ (Ba-122) /Fe bilayers.}
\label{tab:depbedingungen}
\centering
\begin{tabular}{ccccccccc}\hline
\multirow{3}{*}{\textbf{{\textit x$_\mathrm{nom}$ }}}&\multirow{3}{*}{\textbf{{\textit x$_\mathrm{AES}$}}} & \multirow{3}{*}{\textbf{dep.-temp.}}&\multicolumn{2}{c}{\textbf{energy density}}&\multicolumn{2}{c}{\textbf{$\varphi$-FWHM (\textdegree)}}&\multirow{3}{*}{\textbf{{\textit c}-axis}}&\multirow{3}{*}{\textbf{$T_\mathrm{c,90}$}}\\
\cline{4-7}
&&&Ba-122&Fe&Ba-122&Fe&&\\
&&(\textcelsius)&(J/cm$^2$)&(J/cm$^2$)&(103)&(110)&(\AA)&(K)\\
\hline
0&0&650&2.45&2.85&1.82&1.23& 12.99775 &\\
0.02&0.015&700&2.50&3.10&1.23&1.04& 12.93307 &13.0\\
0.04&0.035&750&2.65&3.30&0.78&0.75& 12.79980 &27.9\\
0.06&0.049&750&2.75&3.05&0.81&0.87& 12.85912 &27.5\\
0.08&0.075&750&2.50&2.60&0.81&0.80& 12.84935 &24.7\\
0.10&0.107&750&2.75&3.15&0.99&1.11& 12.82454 &20.6\\
0.15&0.134&800&2.75&3.15&1.14&0.81& 12.74101 &22.1\\
0.15&0.132&750&2.50&2.50&1.27&0.87& 12.78704 &20.3\\
\hline
\end{tabular}
\end{table*}

\section{\label{sec:level3}Measurements and Results}

\begin{figure}[htbp]
\centering
\includegraphics[width=\columnwidth]{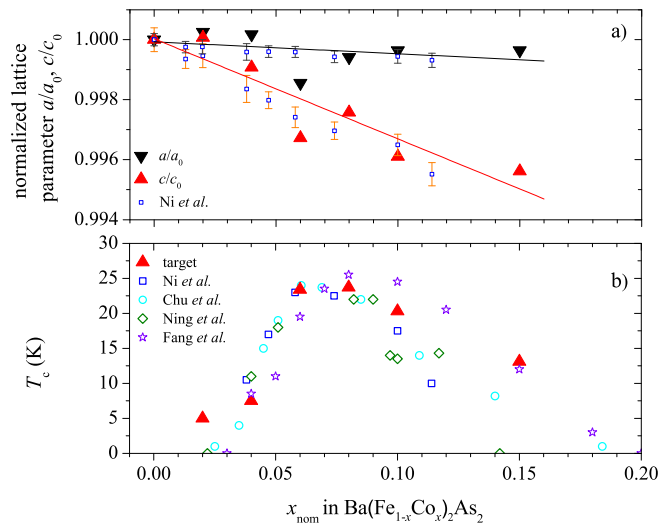}
\caption[lattice parameters of the targets in comparison to the single crystal data from Ni et al.]{(a) {\textit c}-axis and {\textit a}-axis values of the Ba(Fe$_{1-x}$Co$_{x}$)$_2$As$_2$ target powders in dependence of the nominal doping level normalized to {\textit x} = 0 ({\textit a$_0$} = 3.96093 \AA; {\textit c$_0$} = 13.01410 \AA). Linear fits are indicated by lines. The results are compared to the single crystal data from Ni {\itshape et al.}\cite{ni08} (small open squares). The normalized lattice parameters of the Ba(Fe$_{0.94}$Co$_{0.06}$)$_2$As$_2$ target deviate from the linear slope. (b) Critical temperature of the target material in dependence of the Co content, compared to literature data.\cite{ni08, chu09, fang2009, ning09}}
\label{achsetarget}
\end{figure}

The phase purity of the target material was analyzed by powder X-ray diffraction (XRD) in Bragg-Brentano geometry (CoK$_{\alpha}$ radiation with $\lambda$ = 1.7889~\AA). Powder XRD scans of the targets and a following Riet\-veld analysis \cite{rietveld90,rodi} show high phase purity of the target material (the {\textit S}-values for the Rietveld analysis are less than 1.3). The obtained lattice parameters follow the trend of Vegard's law and are similar to results from single crystal measurements (Fig. \ref{achsetarget}(a)).\\
\indent The transition temperatures, {\textit T$_\mathrm{c}$}, of the processed targets were determined by SQUID and VSM measurements. The resulting phase diagram shows a maximum {\textit T$_\mathrm{c}$} of 23.7~K at a nominal composition $x_\mathrm{nom}$ = 0.08 (Fig. \ref{achsetarget}(b)).


Epitaxial growth of the thin films was confirmed by X-ray $\theta$-2$\theta$ scans and texture measurements as shown exemplarily in appendix B for the film grown from the Ba(Fe$_{0.92}$Co$_{0.08}$)$_2$As$_2$ target (Fig. \ref{phiscanb} and Fig. \ref{phiscana}). The epitaxial relation was confirmed to be (001)[100]Ba-122$||$(001)[110]Fe$||$(001)[100]MgO .\cite{Thersleff2010}

\begin{figure}[htbp]
\centering
\includegraphics[width=\columnwidth]{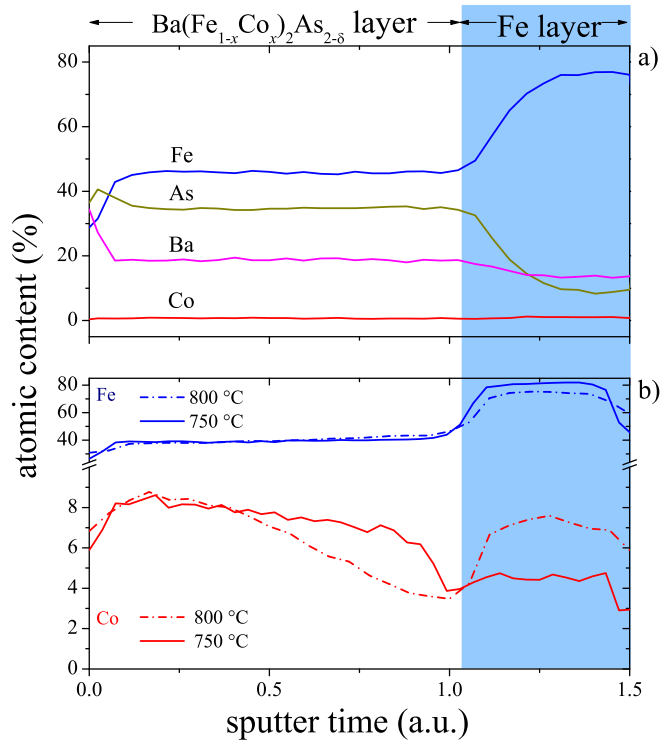}
\captionof{figure}{(a) AES result of the Ba(Fe$_{0.98}$Co$_{0.02}$)$_2$As$_{2-\delta}$ thin film. The As deficiency in the Ba(Fe$_{0.98}$Co$_{0.02}$)$_2$As$_{2-\delta}$ layer shows up to $\delta$ = 0.15 (b) Co dopant and Fe concentrations in the nominal Ba(Fe$_{0.85}$Co$_{0.15}$)$_2$As$_2$ and the Fe layer for different deposition temperatures, 750~\textcelsius\ (solid lines) and 800~\textcelsius\ (dash-dotted lines).}
\label{tcerklaerung}
\end{figure}

\begin{figure}[htbp]
\centering
\includegraphics[width=0.7\columnwidth]{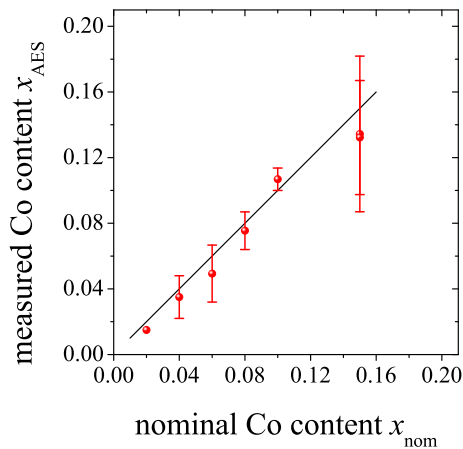}
\captionof{figure}{The doping {\textit x}$_\mathrm{AES}$ measured by AES in the thin films compared to the nominal Co content {\textit x}$_\mathrm{nom}$.}
\label{nomoverAES}

\end{figure}

\begin{figure}[htbp]
\centering
\includegraphics[width=\columnwidth]{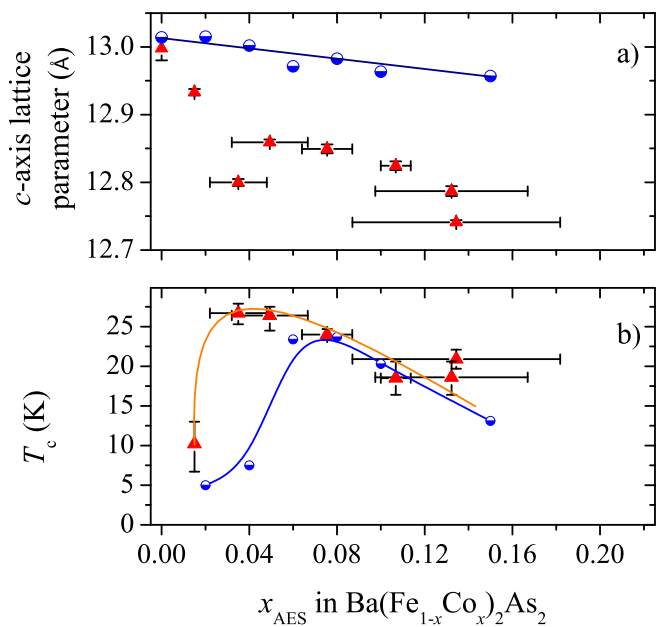}
\captionof{figure}{(a) {\textit c}-axis values for thin films with different Co content, {\textit x}. (b) The phase diagram of the Co doped BaFe$_2$As$_2$ thin films (red). The error in {\textit T$_\mathrm{c}$} is given by the values for {\textit T$_\mathrm{c,90}$} and {\textit T$_\mathrm{c,10}$} and hence the transition width. The error bars for the Co doping represent the Co gradient measured by AES. The blue dots represent the data points of the bulk material, used for PLD.}
\label{phasendiagramm}
\end{figure}

The thickness of the Fe buffer and the Ba(Fe$_{1-x}$Co$_x$)$_2$As$_2$ layer was confirmed on a Focused-Ion-Beam (FIB)-cut cross section in a scanning electron microscope equipped with a FIB stage.\cite{iida:172507} For all films the Fe buffer thickness is between 15 and 20~nm, whereas the Ba(Fe$_{1-x}$Co$_x$)$_2$As$_2$ layer thickness is about 100~nm.
The {\textit T$_\mathrm{c}$} of the samples was determined by the Van der Pauw method using a Physical Property Measurement System (PPMS). A heating rate of  1 K/min and a constant current of 100 $\mu$A were used in the electrical transport measurements. Due to the higher conductivity of the Fe buffer layer and the resulting shunting of the current,\cite{tro2012} the SDW anomaly in the underdoped films cannot be detected through R(T) measurements.
In addition, the $T_\mathrm{c}$ criterion (except for $T_\mathrm{c,0}$) shifts towards lower resistivities and thus to lower temperatures because of the Fe buffer layer. In order to obtain the transition temperatures of the Ba(Fe$_{1-x}$Co$_x$)$_2$As$_2$ layer, the R(T) data were corrected using the method given by S. Trommler {\itshape et al.} \cite{tro2012}\\
\indent A key question in the PLD process addresses the stoi\-chio\-metric transfer of the material from the target to the substrate.
Thin film stoichiometry was determined by Auger electron spectroscopy (AES) where parts of the film were sputtered and the target material was used as a standard. The resulting depth profiles of the films do not only reveal the stoichiometry but also inhomogeneities in the film compositions (Fig. \ref{tcerklaerung}(b)).
The quantification of the film stoichiometry is based on calibrated AES measurement of target material, that was grinded and sputter cleaned using the same conditions as in the depth profiles. Thus surface contamination is removed, preferential sputtering and some peak interferences (signals of Ba partly overlapped by Fe) are considered in the concentration quantification of the layers. The Ba-122 target material with a doping of {\it x$_\mathrm{nom}$} = 0.1 assumed to be in nominal stoichiometry was used to create a set of sensitivity factors for calibration, confirmed also for the {\it x$_\mathrm{nom}$} = 0.02 target material. The calibration thus works well in the Ba(Fe$_{1-x}$Co$_x$)$_2$As$_2$ layer of the films, but cannot be used for quantification in the Fe buffer layer.\\
\indent A remarkable observation is the As deficiency especially in underdoped films compared to the target material. The maximum As deficiency is estimated from AES to be about $\delta$ = 0.15 in Ba(Fe$_{1-x}$Co$_{x}$)$_2$As$_{2-\delta}$ (Fig. \ref{tcerklaerung}(a)). We assume As diffusion into the Fe buffer during film growth. Additionally, diffusion of Co into the Fe buffer layer is noticeable, and its amount changes with varying Co content and deposition temperatures. In the higher doped samples with a nominal Co content of {\textit x}$_\mathrm{nom}$ = 0.15 a large Co gradient over the layer thickness was found for increased deposition temperatures. A diffusion of Co into the Fe buffer layer takes place especially at elevated temperatures where the Co content in the Fe buffer increases. The Co content in Ba(Fe$_{1-{\textit x}}$Co$_{\textit x}$)$_2$As$_2$ thin films deposited at higher temperatures ranges between {\textit x} = 0.20 and {\textit x} = 0.09. A steep Co concentration gradient along the film thickness as well as a strong Co diffusion into the Fe buffer can be observed. In contrast, the Co gradient of the thin film and the Co diffusion into the Fe buffer deposited at 750~\textcelsius\ is significantly smaller (Fig. \ref{tcerklaerung}(b)).
Despite local inhomogeneities, the measured Co concentration, {\textit x$_\mathrm{AES}$}, in the thin films follows the nominal Co content {\textit x$_\mathrm{nom}$} quite well (Fig. \ref{nomoverAES}).\\
\indent The {\textit c}-axis lattice parameter of the Ba(Fe$_{1-x}$Co$_x$)$_2$As$_2$ phase decreases with increasing {\textit x$_\mathrm{AES}$} (Fig. \ref{phasendiagramm} (a)). The {\textit c}-axis of the thin films is always smaller compared to the {\textit c}-axis of the bulk targets. The resulting electronic phase diagram determined from the thin films is shown in Fig. \ref{phasendiagramm}(b). The error bars in {\textit x$_\mathrm{AES}$} indicate the difference between the highest and the lowest doping value determined by AES. The error bars in $T_\mathrm{c}$ show the transition width of the corrected measurements from $T_\mathrm{c,90}$ to $T_\mathrm{c,10}$ ($T_\mathrm{c,90}$ and the $T_\mathrm{c,10}$ being taken at 90~\% and 10~\% of the normal state resistivity at 35~K).

\section{Discussion}
The measured Fe/Co ratio indicates a qualitative agreement between nominal and measured concentrations (Fig. \ref{nomoverAES}). In all thin films, As deficiency was observed with approximately $\delta$ = 0.25 (Fig. \ref{tcerklaerung}(a)). Quantitatively, the stoichiometric transfer from the target to the thin film cannot be exactly determined from our analysis yet.\\
\indent Simultaneously, we have observed a Co concentration gradient along the film cross section for films with a high Co doping level. At elevated deposition temperatures, Co easily diffuses into the Fe buffer. The resulting gradient of the Co dopant in BaFe$_2$As$_2$ is responsible for a sequence of differently doped regions along the thickness. Instead of a homogeneously overdoped film, there are also optimally Co doped layers which may be responsible for a higher {\textit T$_\mathrm{c}$} compared to bulk values in the electronic phase diagram.\\
\indent The explanation for the enhanced {\textit T$_\mathrm{c}$} values in underdoped films is more subtle. The Co concentration gradient is less pronounced (Fig. \ref{tcerklaerung}(a)), hence, it cannot be considered as primary candidate for the high {\textit T$_\mathrm{c}$} values. Strain effects cannot be excluded completely, however, a quantification is still missing. In addition, the role of the Fe buffer layer between substrate and iron pnictide thin film is yet unclear. First, the buffer layer is supposed to act as strain absorber. Second, it has to be further investigated how its ferromagnetism influences the SDW and/or the superconducting state of BaFe$_2$As$_2$ due to proximity. \\
\indent Instead of strain we propose disorder as primary candidate to explain the observed {\textit T$_\mathrm{c}$} enhancement. Impurities in superconductors (one aspect of disorder) are generally classified according to their scattering potential as magnetic or non-magnetic and act as Cooper pair breaker leading to a {\textit T$_\mathrm{c}$} suppression described by the Abrikosov-Gor'kov formula in conventional superconductors.\cite{abri1961} The situation is more complex in multiband superconductors, where the effects of impurity scattering are dependent on the scattering strength and the anisotropy of the scattering potential, the coupling constant matrix, the symmetry of the superconducting gap (s$^{\pm}$, s$^{++}$, d-wave), and, finally the scattering channels commonly denoted as intraband and interband.\cite{PhysRevB.55.15146}\\
\indent In the case of underdoped Ba(Fe$_{1-{\textit x}}$Co$_{\textit x}$)$_2$As$_2$, superconductivity coexists microscopically with a SDW state, {\itshape i.e.} SDW and superconductivity compete for the same electrons.\cite{PhysRevLett.103.087002, pratt09} Shifting the superconducting dome in the electronic phase diagram is thus a result of any mechanism suppressing the SDW but leaving the Cooper pairing intact.\\
\indent A very recent theoretical work proposes that non-magnetic impurity scattering suppresses the SDW transition more strongly than the superconducting transition. Thus it is responsible for an enhancement of {\itshape T$_\mathrm{c}$} in underdoped Ba(Fe$_{1-{\textit x}}$Co$_{\textit x}$)$_2$As$_2$.\cite{2012arXiv1203.3012F} This was experimentally demonstrated by additional Cu doping of Ba(Fe$_{1-{\textit x}}$Co$_{\textit x}$)$_2$As$_2$.\cite{PhysRevB.82.024519} Under the assumption of s$^{\pm}$ pairing\cite{mazin08, Kuroki2008} and a dominant intraband impurity scattering, the superconducting state survives but at the same time the SDW state is weakened.\cite{2012arXiv1203.3012F}\\
\indent The same argument, however, holds for any additional (non-magnetic or magnetic) intraband scattering in the hole pocket\cite{PhysRevB.55.15146, chubukov08, 0295-5075-85-4-47008} that acts detrimentally on the SDW formation without strong influence on (s$^{++}$ or s$^{\pm}$) superconductivity. This might be the case for the thin films investigated here, where we suggest As vacancies acting as active impurities. Although As vacancies may be primarily regarded as non-magnetic defects, they are also able to form localized magnetic moments.\cite{Kikoin2012, kikoin2} Similarily, Se vacancies have been reported to act as magnetic defects in FeSe$_\mathrm{1-\delta}$.\cite{Lee2008} Investigations of As deficient oxy\-pnic\-tides found a ferromagnetically ordered Fe cluster around As vacancies\cite{fuchs08, fuchs09, vadim11, hammerath10} which does not suppress {\textit T}$_\mathrm{c}$. A magnetic impurity scattering contribution due to As vacancies would favour s$^{\pm}$ pairing in the Ba(Fe$_{1-{\textit x}}$Co$_{\textit x}$)$_2$As$_2$ thin films. However, in the {\itshape dirty limit} a crossover from s$^{\pm}$ to s$^{++}$ is also possible. \\
\indent In addition, Cvetkovic and Tesanovic\cite{0295-5075-85-3-37002} have argued that the itinerant character of the FeAs tetrahedral sublattice is reduced with increased flattening. 
The {\itshape c}-axis lattice constants of Ba(Fe$_{1-{\textit x}}$Co$_{\textit x}$)$_2$As$_2$ thin films are drastically smaller compared to the bulk samples. This structural modification might be responsible for a further stabilization of Cooper pairing in the underdoped regime, however, more systematic results are required in order to qualify and quantify strain effects.

\section{\label{sec:level5}Conclusions}

To summarize, we have studied the superconducting transition in the electronic phase diagram of Co doped BaFe$_2$As$_2$ thin films prepared by PLD. The thin films allow to study the effect of disorder and superconductivity in the {\itshape dirty limit}. We have reported a significant effect on the electronic phase diagram: the measured {\textit T$_\mathrm{c}$} values are higher compared to single crystal data, and the superconducting dome shifts towards lower Co content. The highest {\textit T$_{c,90}$} of 27.9~K was achieved at {\textit x$_\mathrm{AES}$} = 0.035 (\textpm 0.013).  
Generally, this change in the electronic phase diagram for Co doped BaFe$_2$As$_2$ thin films can qualitatively be understood by taking into account disorder effects and impurity scattering. 
Additional (non-magnetic or magnetic) intraband scattering weakens the SDW but has less effect on the superconducting state. The thin films in this work are As deficient compared to single crystals. We thus propose As vacancies as a possible candidate for additional intraband scattering centers in Co doped BaFe$_2$As$_2$ being responsible for a suppression of the SDW in favour of Cooper pairing. A more detailed analysis of the role of As vacancies as impurity scatterers in Co doped BaFe$_2$As$_2$ is currently under investigation. Strain and the influence of the ferromagnetic Fe buffer layer may have similar effects on SDW and superconductivity. Their contribution is unclear and cannot yet be separated or quantified completely. 
Finally, due to the high complexity of disorder effects and impurity scattering in multiband superconductivity as well as the coexistence of superconductivity with a SDW state, the term {\itshape dirty limit} in the iron pnictides may describe qualitatively quite different scenarios and its meaning for superconducting properties has to be specified carefully.

\begin{acknowledgments}

The authors acknowledge the financial support under the HA 5934/3-1 (SPP 1458) from the German Research Foundation. F. Kurth acknowledges financial support of the EU (Iron-Sea under project No. FP7-283141). The authors thank S.-L. Drechsler, E. Reich, A. Reisner, V. Grinenko, and R. H\"uhne for valuable discussions as well as U. Besold and M. K\"uhnel for technical support. The work was carried out at IFW Dresden. 

\end{acknowledgments}

\appendix

\section{Target Preparation}

\begin{figure}[htbp]
\centering
\includegraphics[width=0.6\columnwidth]{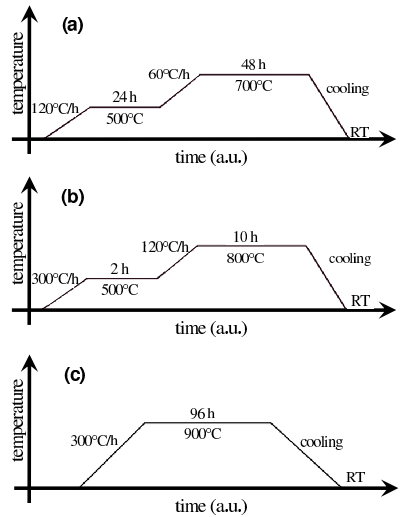}
\caption[heat treatment of the materials]{(a) Heat treatment for BaAs, (b) for Co$_2$As and Fe$_2$As, and (c) for Ba(Fe$_{1-x}$Co$_x$)$_2$As$_2$.}
\label{fig:heat}
\end{figure}

For the preparation of the BaAs precursor a two-zone quartz ampoule was used to guarantee a safe solid state reaction under Ar-shielding gas atmosphere. Prior to this, the ampoule was baked out under vacuum for 10~h at 500~\textcelsius\ to remove contaminations. The heat treatment for the reaction
\begin{align}\mathrm{Ba + As \rightarrow BaAs} \end{align}
is shown in Fig. \ref{fig:heat}(a) and consists of a heating ramp (120~\textcelsius/h), and a holding step at 500~\textcelsius\ for 24 h. Afterwards the heating continued with a rate of 60 \textcelsius/h up to 700~\textcelsius\ above the sublimation point of As. BaAs was held at this temperature for 48~h before cooling down to room temperature with a fast rate of 600~\textcelsius/h.

For the production of Fe$_2$As and Co$_2$As, mechanically milled elements with a homogeneous grain size of 100 $\mu$m were used. Fe and As as well as Co and As were mixed under consideration of oxygen impurities in the ratio of 2:1, respectively.\\
\indent All milled powders were pressed to cylindrically shaped pellets with a diameter of 10 mm by the use of a hydraulic press (Perkin Elmer) with a pressure of 12$\cdot$10$^6$~N/m$^2$. The pellets were placed in a quartz ampoule, and the heat treatment took place under Ar shielding gas. After a heating to 500~\textcelsius\ with a rate of 300~\textcelsius/h the temperature was held for 2 h. A further ramp to 800~\textcelsius\ with a rate of 120~\textcelsius~and a holding time of 10~h followed. The pellets were cooled down to room temperature (Fig. \ref{fig:heat}(b)).\\
\indent Finally, after milling, mixing via the reaction

\begin{equation}\begin{split}\mathrm{BaAs} + (1-x)\cdot \mathrm{Fe_2As} + x\cdot \mathrm{Co_{2}As}\\ \rightarrow \mathrm{Ba(Fe}_{1-x}\mathrm{Co}_x)_2\mathrm{As}_2 \end{split}\end{equation}

and pressing the powder material to pellets with a diameter of 10~mm, Ba(Fe$_{1-x}$Co$_x$)$_2$As$_2$ was synthesized according to the heat treatment given in Fig. \ref{fig:heat}(c).

\section{XRD data for thin films}

\begin{figure}[htbp]
\centering
\includegraphics[width=\columnwidth]{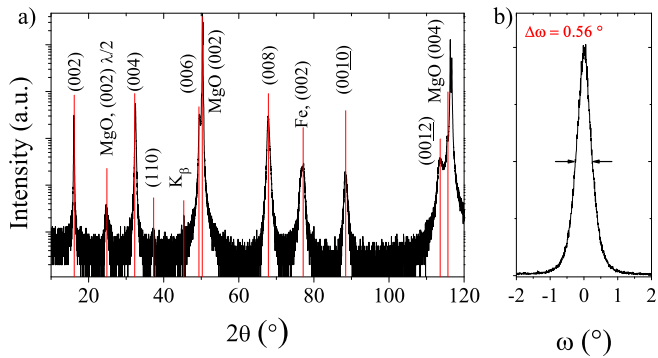}
\captionof{figure}{(a) $\theta$ - 2$\theta$ scan of the thin film grown from a Ba(Fe$_{0.92}$Co$_{0.08}$)$_2$As$_2$ target (b) Rocking curve of the Ba(Fe$_{0.92}$Co$_{0.08}$)$_2$As$_2$ (004) reflection {\textit c}-axis showing the highly {\textit c}-axis textured growth of the film.}
\label{phiscanb}
\end{figure}

\begin{figure}[htbp]
\centering
\includegraphics[width=\columnwidth]{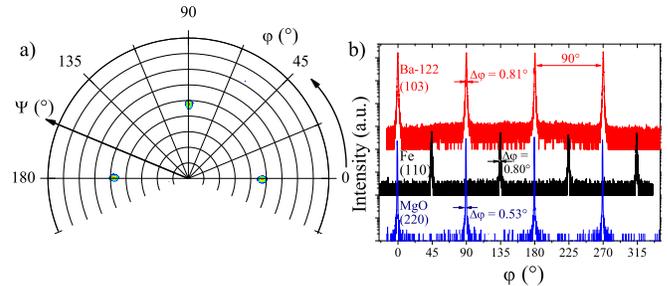}
\captionof{figure}{(a) Pole figure of the Ba(Fe$_{0.92}$Co$_{0.08}$)$_2$As$_2$ (103) reflection, (b) MgO, Fe and Ba(Fe$_{0.92}$Co$_{0.08}$)$_2$As$_2$ $\varphi$-scan.}
\label{phiscana}
\end{figure}

Examplarily the XRD thin film data and the texture data are given by Fig. \ref{phiscanb} and Fig. \ref{phiscana}. The full width at half maximum (FWHM) of the $\varphi$-scans of the Ba(Fe$_{1-x}$Co$_x$)$_2$As$_2$ (103) reflection is between 0.78\textdegree\ and 1.82\textdegree\ (Tab. \ref{tab:depbedingungen}). These values are not corrected for device broadening. Applying the Nelson-Riley extrapolation for $\theta$-2$\theta$ scans, the {\textit c}-axes of the thin films were determined. Within the doping series, the {\textit c}-axis lattice parameters follow Vegard's law (Fig. \ref{phasendiagramm}).

\indent The pole figure of the Ba(Fe$_{0.92}$Co$_{0.08}$)$_2$As$_2$ (103) reflection and the $\varphi$-scans of the Ba(Fe$_{0.92}$Co$_{0.08}$)$_2$As$_2$, the Fe (110) and the MgO (220) reflection are presented in Fig. \ref{phiscana}(a) and \ref{phiscana}(b) and demonstrate the epitaxial relation between film and substrate.


\bibliography{rev}

\end{document}